\documentclass[preprintnumbers,amssymb,amsmath,superscriptaddress,letterpaper,nofootinbib,showpacs,twocolumn,floatfix]{revtex4}[10pt]

\usepackage{graphicx}
\usepackage{dcolumn}
\usepackage{bm}
\usepackage{natbib}
\usepackage{calligra}
\usepackage[T1]{fontenc}
\usepackage{egothic}
\usepackage[T1]{fontenc}
\newfont{\rsfsten}{rsfs10 scaled 1200}
\newfont{\rsfsseven}{rsfs10 scaled 1200}
\newfont{\rsfsfive}{rsfs10 scaled 1200}
\usepackage{epsfig}
\usepackage{units}

\newcommand{\be}{\begin{equation}}
\newcommand{\ee}{\end{equation}}
\newcommand{\bea}{\begin{eqnarray}}
\newcommand{\eea}{\end{eqnarray}}

\def\lsim{\mathrel{\raise.3ex\hbox{$<$\kern-.75em\lower1ex\hbox{$\sim$}}}}
\def\gsim{\mathrel{\raise.3ex\hbox{$>$\kern-.75em\lower1ex\hbox{$\sim$}}}}

\begin{document}

\hspace*{130mm}{\large \tt FERMILAB-14-303-A}
\vskip 0.2in

\title{A Critical Reevaluation of Radio Constraints on Annihilating Dark Matter}


\author{Ilias Cholis}
\affiliation{Fermi National Accelerator Laboratory, Center for Particle Astrophysics, Batavia, IL}
\author{Dan Hooper}
\affiliation{Fermi National Accelerator Laboratory, Center for Particle Astrophysics, Batavia, IL}
\affiliation{University of Chicago, Department of Astronomy and Astrophysics, 5640 S. Ellis Ave., Chicago, IL}
\author{Tim Linden}
\affiliation{University of Chicago, Kavli Institute for Cosmological Physics, Chicago, IL}
\date{\today}

\begin{abstract}
A number of groups have employed radio observations of the Galactic Center to derive stringent constraints on the annihilation cross section of weakly interacting dark matter.  In this letter, we show that electron energy losses in this region are likely to be dominated by inverse Compton scattering on the interstellar radiation field, rather than by synchrotron, considerably relaxing the constraints on the dark matter annihilation cross section compared to previous works. Strong convective winds, which are well motivated by recent observations, may also significantly weaken synchrotron constraints. After taking these factors into account, we find that radio constraints on annihilating dark matter are orders of magnitude less stringent than previously reported, and are generally weaker than those derived from current gamma-ray observations.
\end{abstract}


\maketitle

In addition to gamma rays and neutrinos, dark matter annihilations can produce charged cosmic rays. Electrons and positrons generated in such interactions lose energy via processes including synchrotron, inverse Compton scattering (ICS), ionization and bremsstrahlung, leading to a variety of potentially observable multi-wavelength signals. Of particular interest are the constraints on dark matter annihilation that can be placed by considering radio observations of the innermost region surrounding the Galactic Center~\cite{Gondolo:2000pn,Bertone:2001jv,Aloisio:2004hy,Regis:2008ij,Bertone:2008xr,Bringmann:2009ca,Laha:2012fg,Asano:2012zv,Bringmann:2014lpa}.

The rate at which a cosmic ray electron or positron loses energy via synchrotron and ICS is given by:
\begin{eqnarray}
\frac{dE}{dt} &=& \frac{4}{3} \sigma_T \rho_{\rm mag} \bigg(\frac{E_e}{m_e}\bigg)^2+\frac{4}{3} \sigma_T \rho_{\rm rad} \bigg(\frac{E_e}{m_e}\bigg)^2 \\
&\simeq & 1.02 \times 10^{-16}\, {\rm GeV/s} \, \bigg(\frac{\rho_{\rm mag}+\rho_{\rm rad}}{{\rm eV}/{\rm cm}^3}\bigg) \,\bigg(\frac{E_e}{{\rm GeV}}\bigg)^2, \nonumber
\end{eqnarray}
where $\sigma_T$ is the Thomson cross section,\footnote{The Thomson cross section for ICS is a valid approximation for GeV-scale electrons. In particular, the difference between the limits obtained using the Klein-Nishina and Thomson cross sections is consistently smaller than a few percent.} and $\rho_{\rm mag}$ and $\rho_{\rm rad}$ are the energy densities in the magnetic and radiation fields, respectively. The energy density of the magnetic field is related to its RMS field strength, $\rho_{\rm mag}=B^2/2\mu_0 \approx 2.2 \times 10^4$ eV/cm$^3 \, \times \,(B/{\rm mG})^2$.

Although it has long been argued that large (mG-scale) magnetic fields are likely to be present within the accretion zone around the Milky Way's central supermassive black hole, Sgr A$^*$~\cite{1992ApJ...387L..25M}, it is challenging to observationally constrain the properties of this field. The recent discovery of the magnetar PSR J1745-2900~\cite{Kennea:2013dfa,Mori:2013yda,2013ATel.5040....1E,Shannon:2013hla}, located at a projected distance of 0.12 pc from Sgr A$^*$, has been useful in this respect. In particular, the observed Faraday rotation measure of this object (${\rm RM}\sim 7 \times 10^{4}$ rad/m$^2$), combined with the observed dispersion measure ($\sim 1.8 \times 10^3$ cm$^{-3}$ pc), has been used to obtain a limit of $B \gsim 50\, \mu$G, assuming that all of the electrons along the line-of-sight are located near the Galactic Center~\cite{Shannon:2013hla,Eatough:2013nva}. For comparison, the local magnetic field is generally estimated to be on the order of a few $\mu$G. 

Previous studies of radio constraints on the annihilation of weakly interacting dark matter particles (WIMPs) in the Galactic Center have often neglected energy loss processes other than synchrotron, as well as the effects of diffusion, free streaming, and convection. In other words, they assume that any electrons injected into the central parsec of the Milky Way lose the entirety of their energy to synchrotron before traveling any significant distance or losing any of their energy through other mechanisms. Constraints on annihilating dark matter that are derived under these assumptions will be unrealistically stringent for a number of reasons:
\begin{itemize}
\item{The inner parsecs of the Milky Way are observed to contain extremely high densities of radiation, causing ICS to dominate over synchrotron and other energy loss processes. In particular, in studying $\sim$100 clouds within 5 pc of the Galactic Center, Wolfire {\it et al.} report the presence of a far-ultraviolet radiation field that is consistent with a centralized source with a luminosity of $L \sim (2-3) \times 10^7 L_{\odot}$~\cite{Wolfire:1990zz} (see also Refs.~\cite{Genzel:2010zy,Do:2013sm,Lu:2013sn,2014arXiv1406.7568F}). Such a radiation field is sufficient to dominate cosmic ray electron energy losses for all but the most optimistic magnetic field models. }

\item{A number of recent observations support the existence of strong outflows, which convect cosmic rays away from the Galactic Center. Refs.~\cite{Crocker:2010rd,Crocker:2010qn}, for example, argue in favor of a convective wind with $v_c\sim$ 100-1200 km/s. More recently, the discovery of the Fermi Bubbles provides us with further evidence in favor of a bipolar convective wind, again with a velocity on the order of 100-1000 km/s~\cite{Su:2010qj}.}

\item{Although little is known about cosmic ray diffusion near the Galactic Center, especially on sub-parsec scales, if one adopts a value for the diffusion coefficient that is similar to those adopted in the literature (on the order of $D \sim 10^{26}-10^{27}$ cm$^2$/s for 1-10 GeV electrons~\cite{Chernyakova:2011zz,Linden:2012iv}), cosmic rays random walk with a typical step size on the order of $l_{\rm step} \sim 2 D/c \sim 0.002-0.02$ pc, and thus move travel approximately 0.28 -- 0.87~pc within a single cooling time assuming the ISRF and magnetic field energy density at calculated at 0.12~pc. Inside this regime, where the diffusion constant becomes on the same scale as the region of interest (and diffusion enters a free-streaming limit) would allow electrons injected within the innermost parsec of the Galactic Center to escape the region before losing most of their energy through synchrotron or other processes.}

\end{itemize}

\begin{figure}[!t]
\includegraphics[width=3.40in,angle=0]{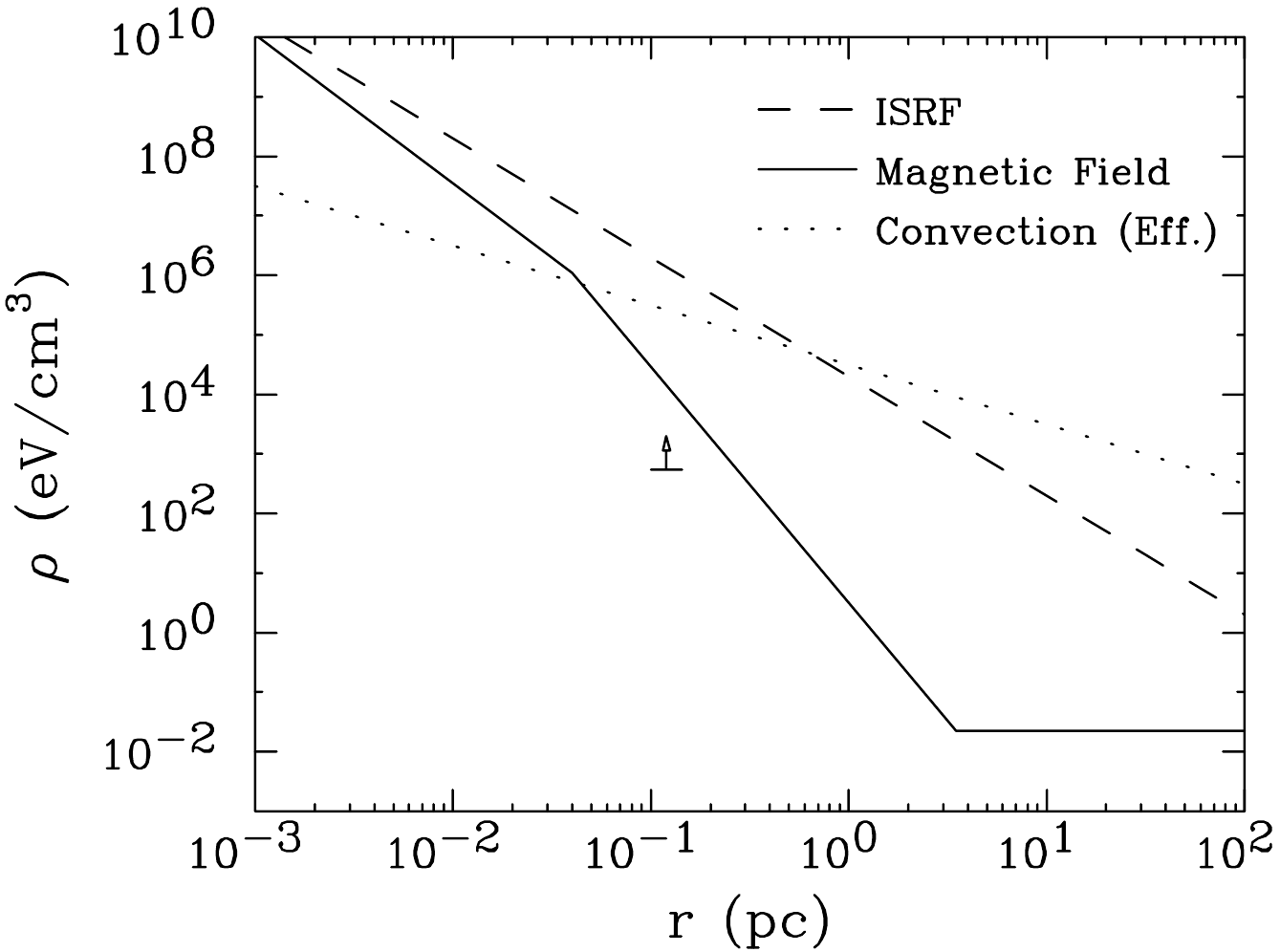}
\caption{The models used in our calculations for the energy density of the magnetic field and of the interstellar radiation field (ISRF) in the region surrounding the Galactic Center. The magnetic field is taken to be near the equipartition value within the accretion zone around Sgr A$^*$ and drops as $B\propto r^{-2}$ outside of that region. We also show the lower limit on the B-field at r=0.12 pc, as derived from recent observations of the magnetar PSR J1745-2900~\cite{Shannon:2013hla,Eatough:2013nva}. For the ISRF, we adopt the profile derived from the results of Ref.~\cite{Wolfire:1990zz}, assuming a centrally located source. The convection line denotes the effective impact of a wind moving cosmic rays away from the Galactic Plane at a velocity of 100 km/s (defined as the energy density in magnetic or radiation fields that would lead to an energy loss time equal to the time required for a 1 GeV electron to convect a distance $r$).}
\label{energydensity}
\end{figure}

In Fig.~\ref{energydensity}, we plot our default model for the energy densities of the magnetic and radiation fields in the region surrounding the Galactic Center.  For the magnetic field, we adopt the profile recently used in Ref.~\cite{Bringmann:2014lpa}:

\begin{equation}
B = 
\begin{cases} 
	7.2~(\frac{0.04~pc}{r})^{5/4}~\mbox{mG}, & r \leq 0.04~pc 
    \\ 7.2~(\frac{0.04~pc}{r})^{2}~\mbox{mG}, & r > 0.04~pc  
\end{cases}
\end{equation}

The normalization in this model is not far from the equipartition value within the accretion zone, and is consistent with the constraint derived from observations of PSR J1745-2900 (shown as an arrow at $r=$0.12 pc). While we consider this model to be plausible, one should keep in mind that it remains largely unconstrained by observations and at this time remains quite speculative. Notably, the analysis of~\cite{Eatough:2013nva} employs a one-zone Faraday screen model and places a lower-limit of 8~mG on the magnetic field strength at the position of the magnetar PSR J1745-2900 (0.12 pc from the galactic center). This limit is significantly higher than the 50~$\mu$G lower-limit placed by~\cite{Shannon:2013hla}, and exceeds the above magnetic field model by nearly an order of magnitude in this region of space. However, it is difficult to simply renormalize Equation 2 to fit the limit calculated by~\cite{Eatough:2013nva}, as a simple extrapolation of this model would predict a magnetic field which greatly exceeds the 40~G upper limit on the magnetic field strength at the surface of Sgr A*~\citep{2006A&A...450..535E}.

The interstellar radiation field (ISRF) model shown has been derived directly from the results of Ref.~\cite{Wolfire:1990zz}, assuming that the radiation originates from a centrally located source.\footnote{More correctly, the ISRF can be computed by taking into account the radial distribution of young and old stars around the galactic center. Young stars are modeled via a distribution that falls along the line of sight as R$^{-0.93}$, while old stars are modeled with a distribution that falls along the line of sight as R$^{-0.16}$~\citep{Do:2013sm}. We find that this has a negligible effect on our results, decreasing the energy density of the ISRF by a factor of 3 at 0.01~pc, but the ISRF energy density by a factor of 5 at 1~pc.} The comparative strength of these curves indicates the relative fraction of electron energy which will produce either $\gamma$-ray emission via the ICS of the ISRF, or radio emission through synchrotron radiation in the galactic magnetic field, allowing us to estimate the maximum radio signal which may plausibly result from dark matter annihilation in scenarios where cosmic-ray diffusion is ineffective at transporting electrons away from the galactic center region. Additionally, We plot in this figure a curve representing the effectiveness of a 100 km/s convective wind at removing electrons at a given radius to a distance twice as far from the galactic center. This effective energy loss-rate is normalized to the above curves assuming an electron energy of 1~GeV.

\begin{figure*}[!t]
\includegraphics[width=3.40in,angle=0]{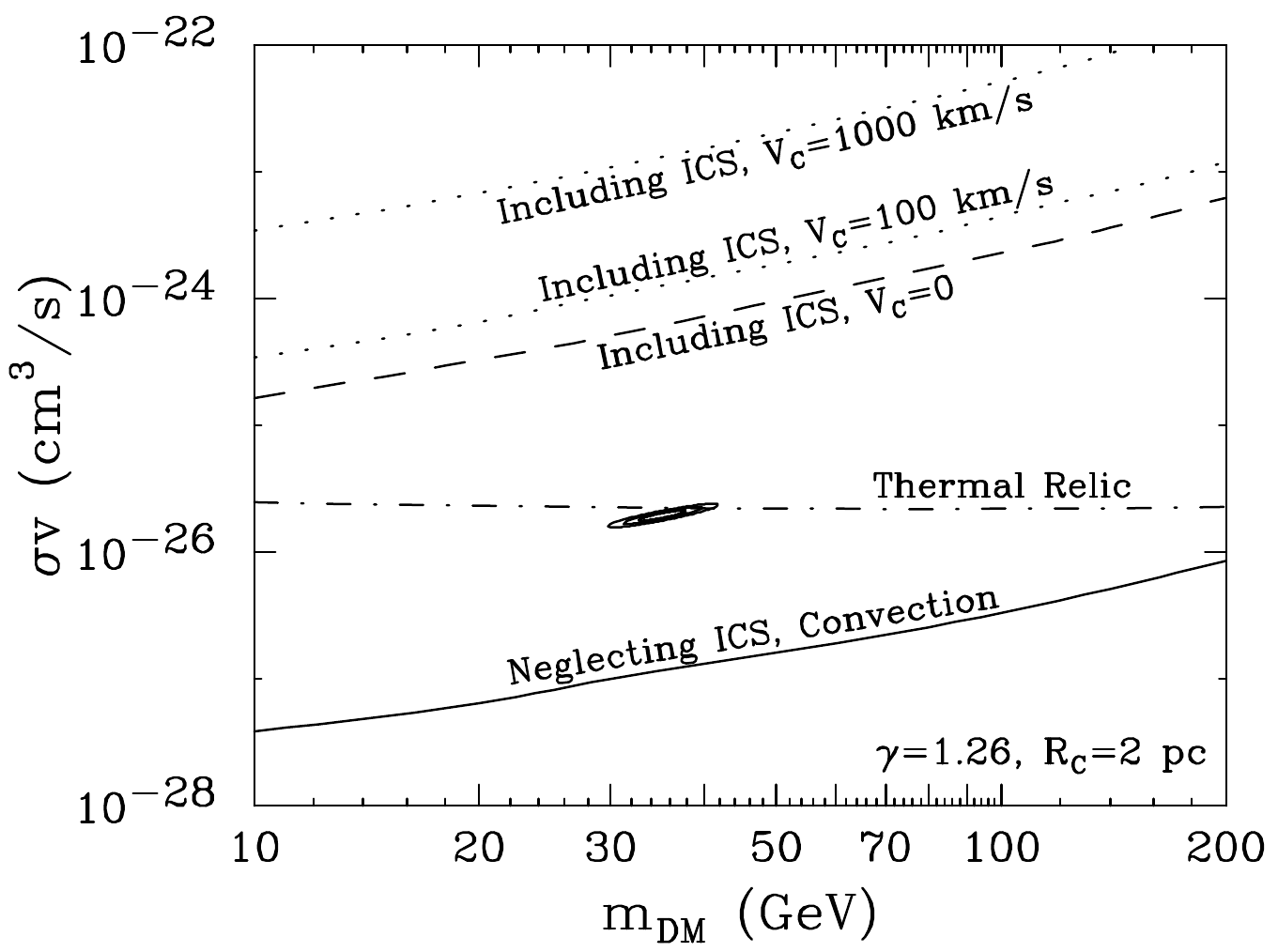}
\includegraphics[width=3.40in,angle=0]{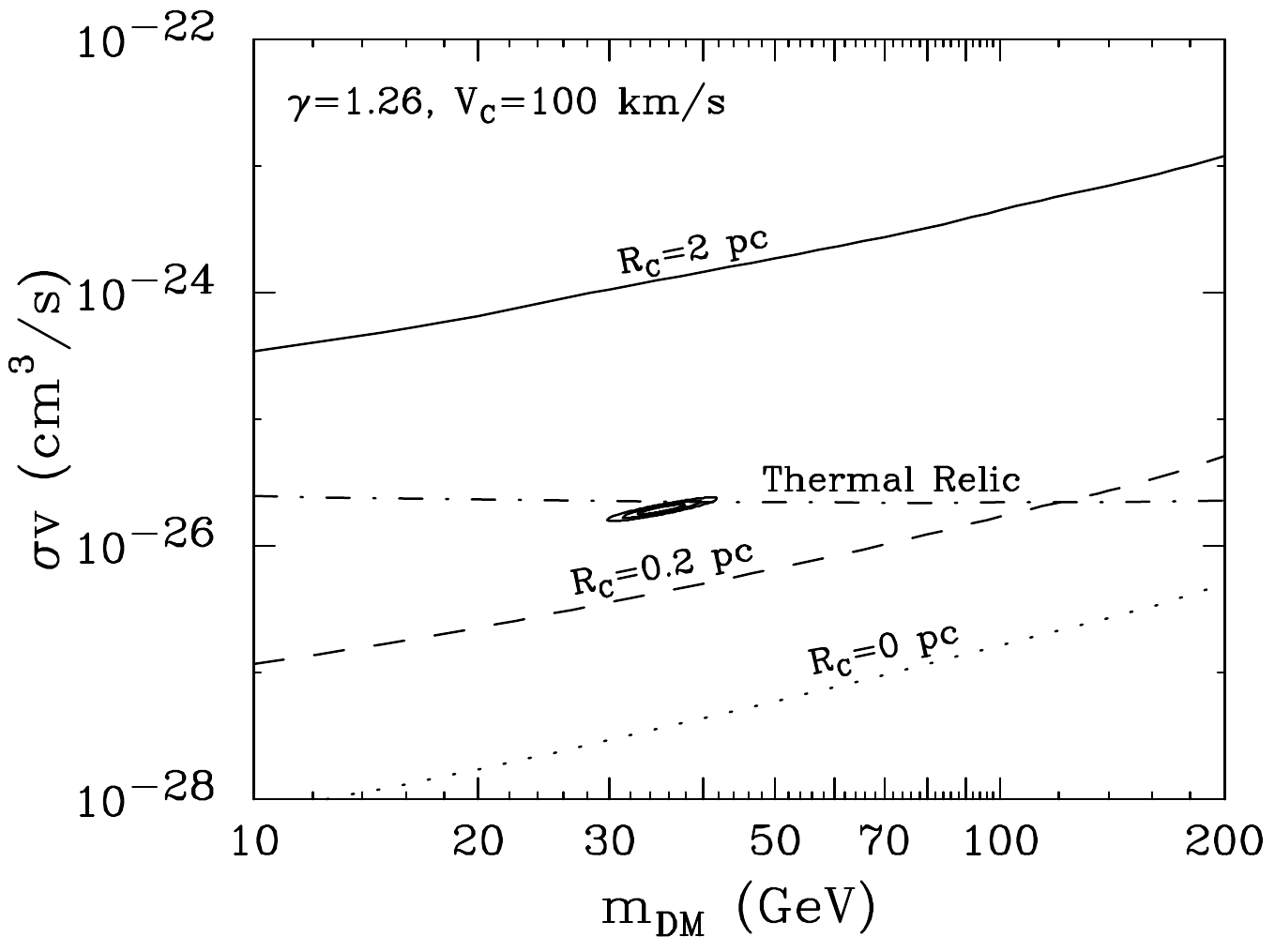}
\caption{Constraints on the dark matter annihilation cross section (to $b\bar{b}$) from 408 MHz radio observations of the central 0.04 arcseconds around Sgr A$^*$. In the left frame, the solid curve neglects both inverse Compton scattering (ICS) and convection, as is often assumed in the literature. The dashed and dotted curves represents the same limit, but including ICS and/or convection. In each case, we have adopted the magnetic field and ISRF models shown in Fig.~\ref{energydensity}  and a dark matter distribution which follows a generalized NFW profile with an inner slope of $\gamma=1.26$. In the left frame, we assume that the dark matter density is flat within a core radius of 2 pc, whereas in the right frame we show results for three different choices of core radius, $R_c=$2.0, 0.2 and 0 pc. For comparison, we also show as closed contours the region favored by the analysis of Fermi data by Daylan {\it et al.}~\cite{Daylan:2014rsa}.}
\label{limits1pt26}
\end{figure*}

To derive constraints on the dark matter annihilation cross section, we make use of radio observations from the Very Large Array at 330~MHz, which limit the  maximum flux density coincident with the position of Sgr~A* to be 80~$\pm$~15~mJy~beam$^{-1}$ with a beam size of 6".8~$\times$~10".9~\citep{2004ApJ...601L..51N}.\footnote{For non-radio astronomers, a Jansky (Jy) is a unit of spectral flux density equivalent to $10^{-23}$ erg/cm$^2$/s/Hz.} We utilize a 2$\sigma$ upper limit on this flux in order to set our limits on dark matter annihilation. This limit remains somewhat conservative (for our analysis), since the limits could be additionally weakened by radio absorption in the galactic center. The analysis of \citep{2004ApJ...601L..51N} does not measure any significant radio absorption, and sets a limit on the optical depth $\tau_{330 MHz}$~$<$~0.4, a value which could conceivably weaken the limits expressed above by an additional factor of 1.5.

We note that previous groups (including our own) have typically employed observations at 408~MHz using the Jodrell Bank telescope, which appeared to limit the flux from the inner 4" cone around Sgr A* to $\lsim$~50 mJy~\cite{1976MNRAS.177..319D}. However, it is noted in~\citep{2004ApJ...601L..51N} that these limits failed to take into account the radio dispersion from the free electron population occupying the line of sight between the galactic center and the solar position. Since this disperses any 408~MHz radio signal over a region larger than the interferometric resolution of the Jodrell Bank telescope, the analysis of~\cite{1976MNRAS.177..319D} is highly insensitive to dark matter annihilation signals. The corrected upper limit for dark matter annihilation from the analysis of \cite{1976MNRAS.177..319D} lies at 0.9~Jy~beam$^{-1}$. which lies an order of magnitude above the flux of Sgr A* observed by~\citep{2004ApJ...601L..51N}. Additionally, several groups have also set dark matter constraints using radio data at other frequencies (such as in Refs.~\cite{Bertone:2008xr} and~\cite{Laha:2012fg}, which make use of observations at 5~$\times$~10$^{4}$ GHz~\cite{Genzel:2003as} and lower-resolution 330 MHz observations~\cite{2010Natur.463...65C}, respectively). However these limits are somewhat less stringent than those from~\citep{2004ApJ...601L..51N}.

In the left frame of Fig.~\ref{limits1pt26}, the solid curve represents the upper limit on the dark matter annihilation cross section (to $b\bar{b}$) derived under the default assumptions adopted in Ref.~\cite{Bringmann:2014lpa}. In particular, this result assumes a dark matter distribution that follows a generalized NFW profile with an inner slope of $\gamma=1.26$, a scale radius of 20 kpc, a local density of 0.3 GeV/cm$^3$, and a flat density core of $R_c=2$ pc. The synchrotron flux from dark matter is then compared to the upper limit from~\citep{2004ApJ...601L..51N}. We use an injected electron spectrum as calculated using PYTHIA~\cite{Sjostrand:2006za},\footnote{By using PYTHIA, we are able to compare our results directly to those from previous groups. Electroweak corrections (as implemented in PPPC~\cite{Cirelli:2010xx}, for example) can impact the resulting limits at a level of up to $\sim$20\%.} and adopt the monoenergetic approximation for synchrotron emission, $\nu=4.7 \, {\rm GHz} \times  (E_e/{\rm GeV})^2 \, (B/{\rm mG})$. Under these assumptions (and neglecting ICS, convection, and diffusion/free-streaming), the resulting limits are indeed very stringent, ruling out simple thermal relics with masses up to a few hundred GeV. When the impact of ICS is included, however, the constraints are weakened by almost three orders of magnitude. The dashed curve in the same frame illustrates this conclusion.

If a strong convective wind is currently active within the central parsec of the Milky Way, it would also be expected to have significant implications for radio constraints on dark matter annihilation.  In particular, such a wind would expel cosmic ray electrons from the Galactic Center before they lose most of their energy to synchrotron or ICS, reducing the predicted flux of radio emission. This is illustrated as the dotted curves in Fig.~\ref{limits1pt26}, for two values of the convection velocity. In Fig.~\ref{energydensity}, we plot an ``effective energy density'' for convection, which is defined as the energy density in magnetic or radiation fields that would lead to an energy loss time, $\tau \equiv E/ (dE/dt)$, for a 1 GeV electron that is equal to the time required to convect across a distance $r$. 
 
The ISRF model used throughout this study is based on the observations of $\sim$10$^2$ gas clouds within 5 pc of the Galactic Center, as reported in Ref.~\cite{Wolfire:1990zz}. More recent observations have shown that the ISRF in the vicinity of the Galactic Center originates from two major sources: an ultraviolet component from a very concentrated population of young stars ($n \propto r^{-1.93}$)~\cite{Genzel:2010zy,Do:2013sm,Lu:2013sn} and a more spatially extended component from older stars ($n \propto r^{-1.16}$)~\cite{2014arXiv1406.7568F,Do:2013sm} (in addition to a subdominant contribution from Sgr A$^*$). Each of these two stellar components contributes a few times $10^{7} L_{\odot}$ within the innermost parsecs of the Galaxy. Given the sum of these observed profiles, we find that the energy density of the ISRF dominates over that of the magnetic field (given the $B$-field model shown in Fig.~\ref{energydensity}) throughout the entire volume of the Galactic Center beyond $\sim$0.01 pc of Sgr $A^*$.

\begin{figure*}[!t]
\includegraphics[width=3.40in,angle=0]{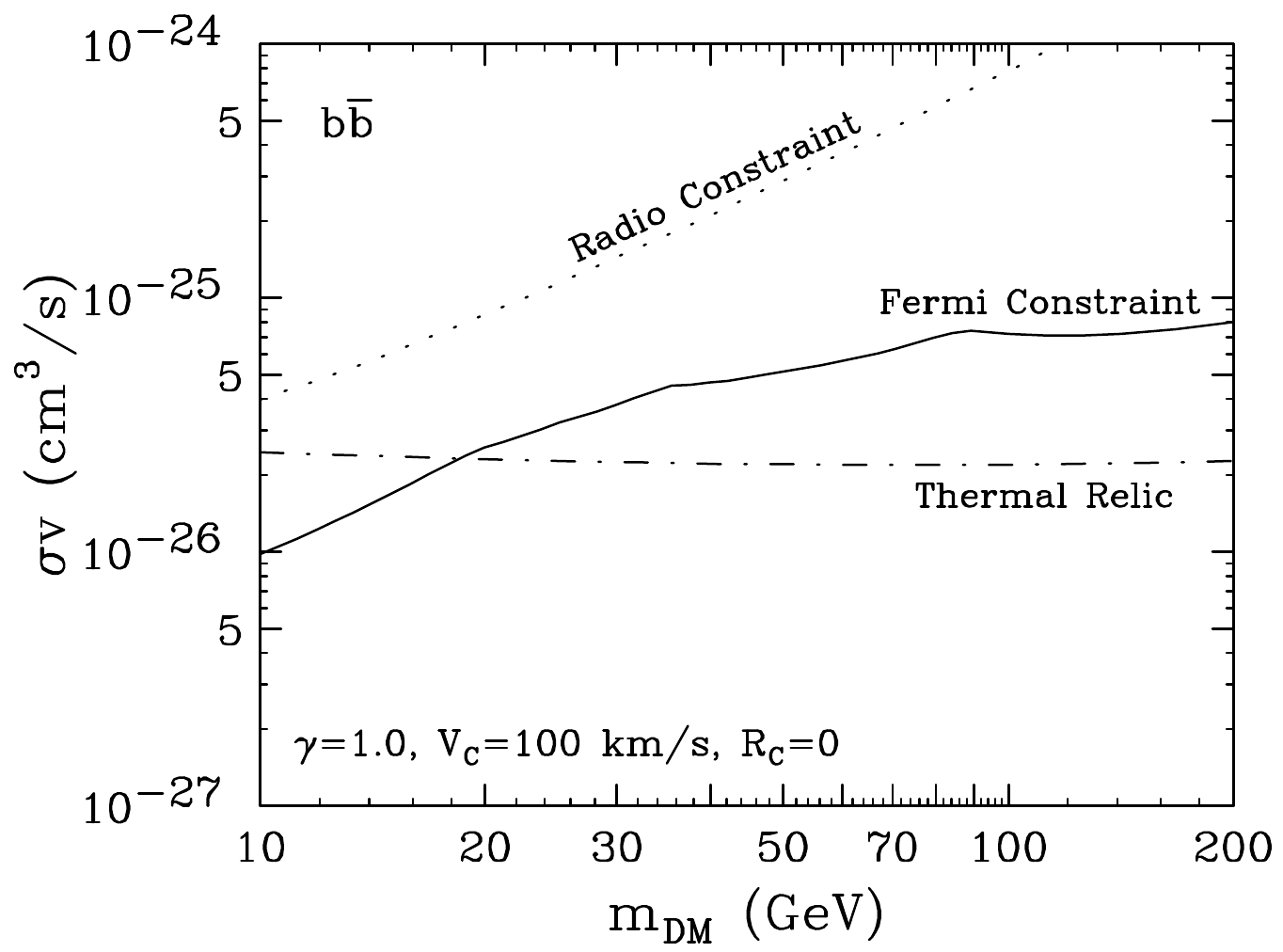}
\includegraphics[width=3.40in,angle=0]{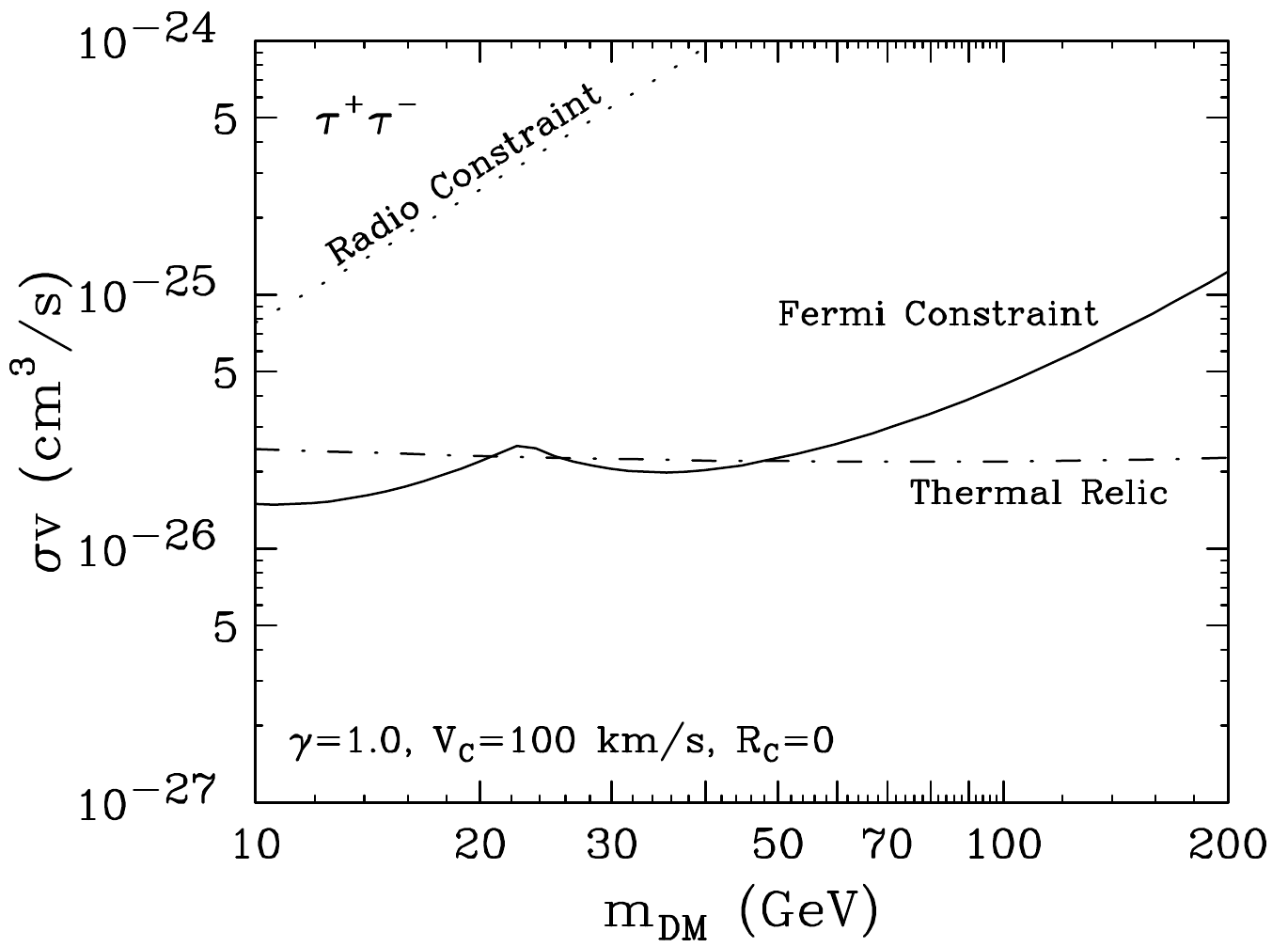}
\caption{A comparison of the constraints derived from radio and gamma-ray observations of the Galactic Center (are reported in Ref.~\cite{Hooper:2012sr}), assuming an NFW profile ($\gamma=1$). Even if one assumes that diffusion/free-streaming can be neglected, and that the dark matter profile and magnetic field models can be accurately extrapolated into the Galactic Center, the resulting radio constraints are generally less stringent than those derived from gamma-ray observations.}
\label{limits}
\end{figure*}

In addition to these observations, there is another line of reasoning that supports the conclusion that cosmic ray electrons in the Galactic Center do not lose most of their energy to synchrotron. The spin-down power of the recently discovered magnetar PSR J1745-2900 is $\dot{E} \approx 2\times 10^{33}$ erg/s $\times \,(B/10^{14}\,{\rm G})^2$. In order for the synchrotron emission from the electrons injected from this source to not exceed the flux observed at 408 MHz, less than 0.2\% of the spin-down power can go into synchrotron.\footnote{In producing this estimate, we have adopted an injected electron spectrum of the form $dN_e/dE_e \propto E_e^{-1.5}$ between 1 and 1000 GeV.} Although this fraction is quite low, it is perhaps not an inconceivable value. The magnetar in question, however, is thought to be only one of a large population of pulsars present within the inner fraction of a parsec around the Galactic Center. In particular, the large number of massive stars and the enhancement in the X-ray binary density observed in the region~\cite{Muno:2005dy} leads one to expect $\sim$100-1000 pulsars to reside within $\sim$0.02 pc of Sgr A$^*$~\cite{Pfahl:2003tf} (see also Refs.~\cite{Dexter:2013xga,Zhang:2014kva,Chennamangalam:2013zja}). The collective synchrotron emission from such a large population of pulsars would almost certainly exceed the radio flux observed from the region unless most of the energy in cosmic ray electrons is not locally emitted as synchrotron.

In the right frame of Fig.~\ref{limits1pt26} we plot limits, including ICS and convection (with $v_c=100$ km/s), for three different choices of the core radius of the dark matter profile. If the dark matter distribution does not continue to rapidly increase as one approaches the innermost parsec around the Galactic Center, radio constraints fall well short of excluding the thermal cross section.

Based on the combination of energy loss mechanisms including ICS,
convection, and diffusion, we find that radio constraints are
competitive with those derived from gamma-ray and other observations
only if {\it all} of the following hold true:
\begin{itemize}
\item{The dark matter density continues to rise (for example as $\rho \propto r^{-1}$) within the innermost parsec of the Galactic Center. As this scale is well below the resolution of numerical simulations, we have little insight into whether this is or is not the case.}
\item{The magnetic fields continue to rise within the innermost parsec, allowing synchrotron to be competitive with energy losses from ICS.}
\item{Cosmic ray electrons must behave diffusively (and not efficiently free-stream) within the central parsec. This would require a low diffusion coefficient, $D \lsim 10^{26}$ cm$^2$/s.}
\end{itemize}

If any of these three criteria are not met, the constraints on dark matter annihilation derived from radio constraints will be very weak. And even if we optimistically assume that the dark matter profile and magnetic field models can be accurately extrapolated into the Galactic Center, and neglect any free-streaming, the resulting constraints are not necessarily more stringent that those derived from gamma-ray and other observations. For example, in Fig.~\ref{limits}, we compare radio constraints to those derived from Fermi observations of the Galactic Center~\cite{Hooper:2012sr}, assuming an NFW profile with a canonical value for the inner slope, $\gamma=1$. For neither annihilations to $b\bar{b}$ or $\tau^+ \tau^-$ do the radio constraints exceed those provided by Fermi.  And although radio observations could provide the most restrictive constraints in more cuspy scenarios ($\gamma >1$), this would only be the case if all three of criteria listed above are satisfied. 

{\it In summary}, we have revisited constraints on annihilating dark matter as derived from radio observations of the Galactic Center.  We find that when inverse Compton scattering with the interstellar radiation field is taken into account, such constraints are weakened by almost three orders of magnitude. If strong convective winds are present in this region (as is supported by recent observations), these constraints will be weakened further. Under relatively optimistic assumptions (regarding magnetic fields, diffusion, {\it and} the dark matter density within the innermost parsec of the Galaxy), radio constraints are comparably stringent to those derived from gamma-ray observations. While there are significant uncertainties in several parameters, most importantly the strength of the galactic center magnetic field, the very reasonable parameter space choices considered in this paper make it difficult to imagine the creation of resilient radio constraints on galactic center dark matter annihilation that fall below the levels presently explored by Fermi.

                  
\smallskip                  
                  
{\it Acknowledgements}: We thank an anonymous referee and Farhad Yusef-Zadeh for comments concerning the application of the limits from~\cite{1976MNRAS.177..319D} to this work. This work has been supported by the US Department of Energy. TL is supported by the National Aeronautics and Space Administration through Einstein Postdoctoral Fellowship Award Number PF3-140110.              
                  
\bibliography{radiocritical}
\bibliographystyle{apsrev}

\end{document}